\documentclass{raa}           
\usepackage{graphicx,times}
\usepackage{epstopdf}
\usepackage{natbib}
\usepackage{amssymb,amsmath}
\bibpunct{(}{)}{;}{a}{}{,}

\usepackage[a4paper=true,dvipdfm=true,pagebackref=true]{hyperref}
\hypersetup{pdftitle = The title of my PDF, pdfauthor = My name, pdfsubject= The subject, pdfkeywords = keyword1 keyword2 keyword3} 
\hypersetup{colorlinks = true, linkcolor = green, anchorcolor = red, citecolor = blue, filecolor = red, pagecolor = red, urlcolor = red}

\begin{document}

   \title{Statistical Improvement in Detection Level of Gravitational Microlensing Events from their Light Curves}

 \volnopage{ {\bf 2016} Vol.\ {\bf X} No. {\bf XX}, 000--000}
   \setcounter{page}{1}

   \author{Ichsan Ibrahim\inst{1,4,6}, Hakim L. Malasan\inst{1,3}, Chatief Kunjaya\inst{1,5},  Anton Timur Jaelani\inst{1,7}, Gerhana Puannandra Putri\inst{1}, Mitra Djamal\inst{2}
   }

   \institute{Department of Astronomy, Institute Technology of Bandung, Bandung, 
Indonesia; {\it ichsan.ibrahim@s.itb.ac.id}\\
        \and
            Department of Physics, Institute Technology of Bandung, Bandung, Indonesia\\
        \and
            Bosscha Astronomical Observatory, Institute Technology of Bandung, Lembang, Indonesia\\
	\and
Department of Informatics, STMIK-Indonesia Mandiri, Bandung, Indonesia\\
\and 
Universitas Ma Chung, Malang, Indonesia\\
\and
Faculty of Shariah, Institut Agama Islam Ternate, Ternate, North Molucca, Indonesia\\ 
\and
Astronomical Institute, Tohoku University, Sendai, Japan\\
\vs \no
   {\small Received 0000 00 00; accepted 0000 00 00}
}

\abstract
{In Astronomy, the brightness of a source is typically expressed in terms of magnitude. Conventionally, the magnitude is defined by the logarithm of the received flux. This relationship is known as the Pogson formula. For received flux with a small signal to noise ratio (S/N), however, the formula gives a large magnitude error. We investigate whether the use of Inverse Hyperbolic Sine function (after this referred to as the Asinh magnitude) in the modified formulae could allow for an alternative calculation of magnitudes for small S/N flux, and whether the new approach is better for representing the brightness of that region. We study the possibility of increasing the detection level of gravitational microlensing using 40 selected microlensing light curves from 2013 and 2014 season and by using the Asinh magnitude. The photometric data of the selected events is obtained from the Observational Gravitational Lensing Experiment (OGLE). We found that the utilization of the Asinh magnitude makes the events brighter compared to using the logarithmic magnitude, with an average of about $3.42 \times10^{-2}$ magnitude and the average of the difference of error between the logarithmic and the Asinh magnitude is about $2.21 \times10^{-2}$ magnitude. The microlensing events, OB 140847 and OB 140885 are found to have the largest difference values among the selected events. Using a Gaussian fit to find the peak for OB140847 and OB140885, we conclude statistically that the Asinh magnitude gives  better mean squared values of the regression and narrower residual histograms than the Pogson magnitude. Based on these results, we also attempt to propose a limit of magnitude value from which the use of the Asinh magnitude is optimal for small S/N data.
}

   \authorrunning{Ibrahim, I. et al. }            
   \titlerunning{Statistical Improvement in Detection Level }  
   \maketitle

%
\section{Introduction}           
\label{sect:intro}
Gravitational lensing is predicted by Einstein's theory of general relativity. The gravitational field of a massive object causes the light rays passing it to bend. The more massive the object, the stronger its gravitational field and hence the greater the bending of light rays. The images may be shifted from their original locations and distorted.  The effects of light deflection due to a gravitational field is called gravitational lensing. Gravitational lensing is divided into three groups (Schneider et al.~\cite{schneider86}): (1). Strong lensing, where there are visible distortions such as the formation of Einstein rings, arcs, and multiple images. (2). Weak lensing, where the distortions of background sources are much smaller and can only be detected by analyzing large numbers of sources in a statistical way to find coherent distortions of only a few percent. The lensing shows up statistically as a preferred stretching of the background objects perpendicular to the direction of the center of the lens. (3). Microlensing, where no distortion in shape can be seen but the amount of light received from a background object changes in time as it passes behind a lens. In one typical case, the stars in The Galaxy may act as the lensing object and a star in the Bulge of Galaxy or a remote galaxy as the background source.

\begin{figure}
   \centering
   \includegraphics[width=15.0cm, angle=0]{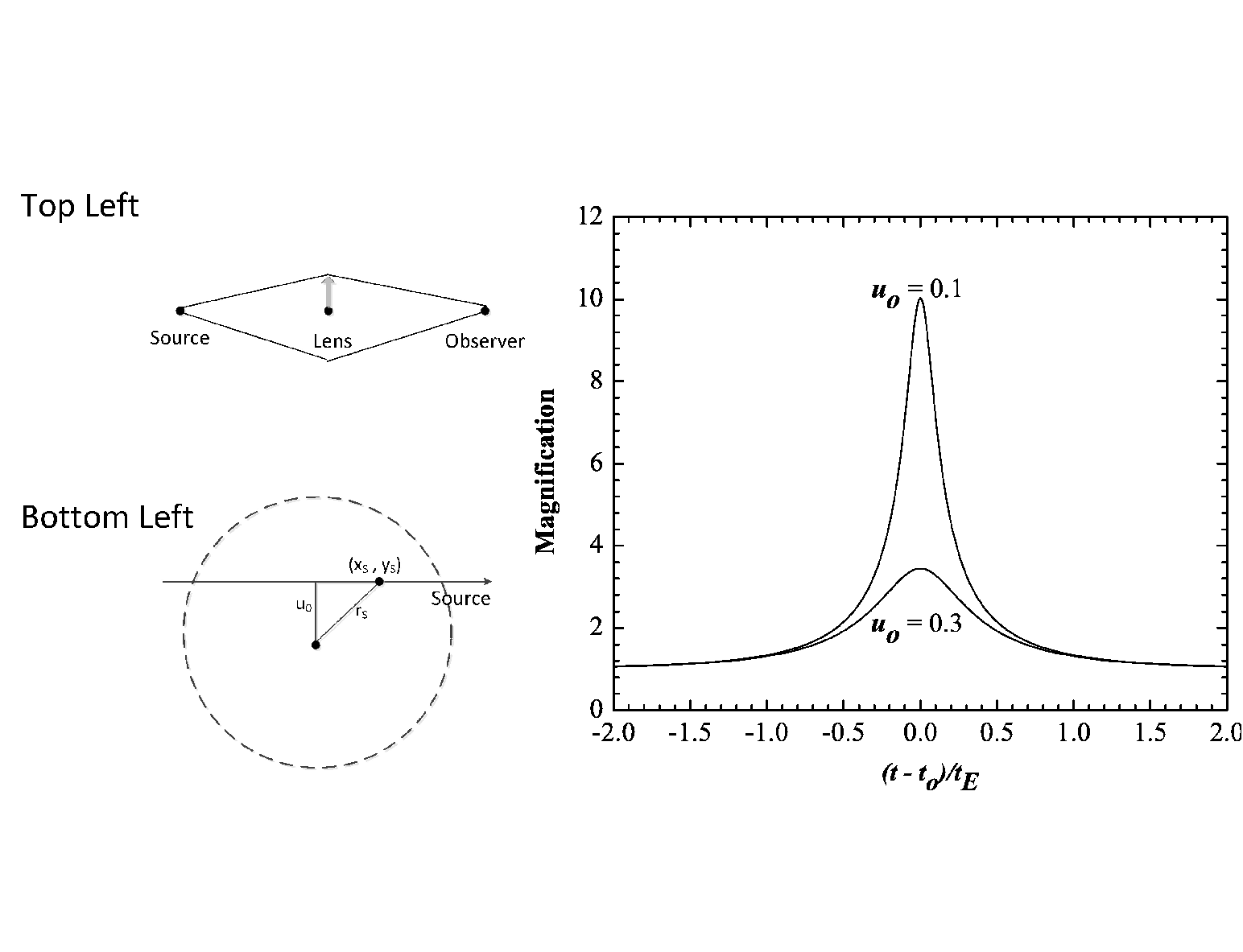}
   \caption{The top left panel shows a side on view of the geometry of gravitational microlensing where a lens moves across the line of  sight towards a background source. The bottom left panel shows an illustration of the lens position and the source trajectory. The position of an observer is in line with the plane of the lens and the source. The lens is at the origin and the source moves across the line of sight along the x-axis. The closest approach of the source ($x_s$ = 0)  is achieved at time $t = t_0$, then $r_s$ is the distance between the lens and the source and the dimensionless source position along the trajectory is given by $x_s = (t $−-$ t_0) / t_E$ and $y_s = u_0$ (also known as the impact parameter). The impact parameter is associated with the gravitational effect of the lens on light from the source which passes near the lens. The right panel shows two light curves associated with two dimensionless impact parameters of lensing events ($u_0$) = 0.1 and 0.3. The time on the horizontal axis is centered on the peak time $t_0$ when there is the closest approach of the source and is normalized to the Einstein radius crossing time $t_E$. (Adaption from Mao~\cite{mao08}).
   }
   \label{Fig1}
   \end{figure}

For star microlensing (from now on termed microlensing), the lensing object is a stellar mass compact object (Mao~\cite{mao08}). The lens is situated on the line of sight from the Earth to a background source (the top left side panel in Figure \ref{Fig1}). A background source radiates light rays passing the lens at different distances and directed towards the lens. The light rays will be bent at a particular bending angle that is affected by their distances to the lens. The bending angle increases with decreasing distance from the lens, and there is a unique distance such that the ray will be deflected just enough to hit the Earth. The unique distance is called the Einstein radius. If the lens moves closer to the line of sight, the effect of gravitation on light rays will be strengthened, so that the source (background) appears brighter. If the lens moves away, the background star brightness will revert to normal. Because of rotational symmetry about the Earth-source axis and when the lens, source and observer are perfectly aligned, an observer on Earth could see the images form a ring (called the Einstein ring). The Einstein ring is centered on the `true' position of the source (the projected position of the source on the plane of the sky). For any other source position, an observer on Earth could see an image, and the source will be mapped into two annuli, one inside the Einstein ring and one outside. In microlensing, the separation angle of the image is to small be visible (smaller than milli arc seconds). We can observe only the changing of the brightness of the source as a function of time, known as the microlensing light curve (after this referred to as the light curve).

The light curve will have a symmetric shape if the motion of the lens, the observer, and the source can be considered as linear movements. The right side panel in Figure~\ref{Fig1} shows two light curves with two impact parameters ($u_0) = 0.1$  and $0.3$. The impact parameter is associated with the gravitational effect of the lens on light from the source which passes close to the lens. To describe the standard light curve and its parameters, the source trajectory is also shown (bottom left side panel in Figure~\ref{Fig1}). We put the lens at the origin and the source moves along the x-axis across the line of sight. The position of the light source along the trajectory is expressed in dimensionless coordinates ($x_s, y_s$). The first coordinate ($x_s)= (t-T_0)/t_E$ and the second ($y_s) = u_0$. The distance between the lens and the source is $r_s$. From that, the impact parameter $u_0$ (in unit time scales ($t_E$) is considered as the closest distance to the lens when the source moves along the trajectory. The peak time ($T_o$) is the time when the light source would be closest to the lens and the time scale ($t_E$) is defined as how long the source takes to traverse the Einstein ring.  $u_0, t_E$, and $T_0$ are the parameters used to model the standard light curve (equation (1)).

\begin{equation}
	A(t) = \frac{r_s(t)^{2}+2}{r_s(t) \times \sqrt{r_s(t)^{2}+4}}
\label{eq:magnification}
\end{equation}
where

$A(t)$ is the magnification of brightness of a source as function of time,

$r_s(t)$ is the distance between the lens and source as function of time:

$r_s(t) = \sqrt{u_0^{2}+ (\frac{t - t_0}{t_E})^{2}}$.

We also note that the variability due to gravitational lensing is achromatic because photons from any wavelength will follow the same propagation path. Revitalization of research into gravitational microlensing was initiated by Paczynski~\cite{paczynski86}. The probability of observing microlensing, when we look towards Galaxy Bulge, is on the order of one in a million (Udalski et al.~\cite{udalski94a}), and in general, microlensing of a particular object is not repeated. Therefore, it is necessary to conduct a regular and continuous survey of areas with a high density of stars or a large number of stars. The Optical Gravitational Lensing Experiment (OGLE) group is one of the research groups routinely providing data and have implemented an early warning detection system for gravitational microlensing events (Udalski et al.~\cite{udalski94b}). 

In general, the results of photometric observations of astronomical objects provide the flux received by an observer on the Earth (after this referred to the flux ($f$)). The flux is converted into magnitude ($m$) using the Pogson equation (after this 
referred to the conventional method).
\begin{equation}
	m = m_0 - 2.5 \times \log f 
\label{eq:Pogson}
\end{equation}
where

$m_0 \equiv 2.5\times \log ⁡f_0$,

$f_0$ is the flux of an object with magnitude 0.0.

The Error of Magnitude $(\Delta ~m)$ is expressed as:
\begin{equation}
	 \Delta~m =|\frac{ -2.5\times\Delta f}{f \times \ln 10}|
\label{eq:errPogson}
\end{equation}
Conversely, the Error of Flux $(\Delta f)$ can be expressed as follows:
\begin{equation}
	 \Delta~f =|\frac{ f \times\Delta m\times \ln 10}{-2.5}| 
\label{eq:errflux}
\end{equation}

The Pogson formula can work well and give a reliable of Error of Magnitude for objects with large flux. Conversely, for a small flux, which usually also means small signal to noise ratio, we will obtain a large and asymmetric Error of Magnitude. The Error of Magnitude is asymmetric because it has a non-Gaussian and skewed distribution. Therefore it is not suitable for defining the magnitude of faint objects. Lupton et al.~\cite{lupton99} proposed a new set of equations to define the brightness of celestial objects from the measured fluxes directly, using the Inverse Hyperbolic Sine function (Asinh) which can behave well on a small signal to noise ratio as follows
\begin{equation}
	\mu = (m_o -2.5 \times \log b') - a\times \sinh^{-1}(\frac{f}{2b'})
\label{eq:Asinh1}
\end{equation}

\begin{equation}
	Var(\mu) = \frac{a^{2}\times\sigma'^{2}}{4\times b'^{2} + f^{2}} \approx \frac{a^{2}\times\sigma'^{2}}{4\times b'^{2}}
\label{eq:Asinh2}
\end{equation}
where 

$\mu$ is the Asinh magnitude,

$Var(\mu)$ is the variance of the Asinh magnitude,

$a \equiv 2.5\times \log e$ = 1.0857 (Pogson ratio) with $e = 2.7182$,

$b' \equiv f_o \times b$, is a softening parameter used to give same result with the Pogson formula for the large flux), 
   
   with $b =\sqrt{ a} \times \sigma =1.042 \times \sigma$, is  arbitrary "softening" constant that determines the flux level at which linear behavior sets in and also as a optimal setting that balances 
   
   two effects: it minimizes the differences between the Pogson and the Asinh for high S/N data and minimizes
   
   the variances for the low flux,

$\sigma' \equiv f_o \times \sigma$, is combination of noise of the flux ($\sigma$) and the flux of an object at magnitude 0.0 ($f_o$),

$f$ is the flux of the object, in this work, we obtained the flux value by using equation (7). 

This definition of magnitude has been applied in the Sloan Digital Sky Survey/SDSS project (York et al.~\cite{york00}). In 2005, the Asinh magnitude formula had used in the study of photometric standard stars at the Bosscha Observatory, Institute Technology of Bandung (Pujijayanti et al.~\cite{puji06}).

\section{Objectives and Methodology}
\label{sect:Obs}
\subsection{Objectives}
Generally, to determine if the use of the Asinh magnitude gives better results, for example, smaller errors, for small signal to noise photometric data and therefore increase the possibility of detecting fainter microlensing events. Specifically, to obtain the magnitude range in which the Asinh magnitude significantly gives a better detection level than the conventional method for the instruments used by the Early Warning System Observational Gravitational Lens Experiment, from now on, called the EWS-OGLE (http://ogle.astrouw.edu.pl/ogle4/ews/ews.html).
\subsection{Methodology}
In this work, we do two things: 
First, we simulate the flux of an object using inputs such as temperature range, size, and distance of the object. In this work, we chose three distinct values of stellar radii, namely 1-3 Solar radii ($R_\odot$) and stellar temperatures ranges from 3000 to 11000 $K$, correlated with stars of spectral classes A to M, observed in the optical region $\lambda = 5500$ \AA. In addition, we also adopt one value of light source distance, D = 8.5 $kpc$. The error of the Flux was derived by assuming a Poisson distribution as the distribution of the photons from the source. After calculating the flux value and the error of flux from the input data, we then calculated the Pogson magnitude ($m$) and the Error of Magnitude($\Delta m $) using equations (2) and (3). Subsequently, we converted the Pogson magnitude ($m$) into the Asinh magnitude ($\mu$) using equation (5), and its corresponding error by taking the square root of its variance (using equation (6)).

Second, we also applied equations (5) and (6) to convert photometric data of microlensing events from the EWS-OGLE database to the Pogson magnitudes ($m$) and the error of magnitudes ($\Delta m $), and after that to the Asinh magnitude ($\mu$ and $\Delta \mu$). As we could not obtain the flux from EWS-OGLE, we used equation (7), together with the information for $I$-band filter from Table A2 in Bessell et al.~\cite{bessell98} Appendix B, and the previously calculated Pogson magnitude ($m$) values to calculate the flux ($f$). After that, we calculate the error of flux ($\Delta f$) using equation (4). As mentioned in Udalski et al.~\cite{udalski15}, the $I$-band filter in OGLE-IV camera that was used to collect the EWS-OGLE data very closely resembles the standard $I$ -band filter. The equation used to calculate the flux is as follows:
\begin{equation}
	M_\lambda = -2.5 \log f_\lambda - 21.1 - z_p
\label{eq:Bessell}
\end{equation}
where

$M_\lambda$ is the $I$-band filter magnitude from EWS-OGLE,

$f_\lambda$ is the flux of the object/events,

$z_p$ is the zero point =  0.444.

We compare the results between the Asinh method and conventional method, and then we will be able to give our best estimate, for the magnitude range, where the Asinh is recommended to be used as a substitute for conventional methods for instruments used in EWS-OGLE.

\section{Data}
\label{sect:Data}
The OGLE group obtained the photometric data for the EWS-OGLE database by using a 1.3 m Warsaw University Reflector with mosaic CCD camera. The $V$- and $I$- band interference filter set for the OGLE-IV camera. The mosaic camera was composed of 32 thin E2V44-82 2048 $\times$ 4096 CCD chips (Udalski et al.~\cite{udalski15}). The reflector was mounted on a Ritchey-Chrétien system at Las Campanas Observatory, Chile $(\phi = 29^0$ 00' $ 36.9'' S;  \lambda = 70^0$ 42' $ 5.1'' W)$. For the purpose of this work, 19 microlensing events were selected from the 2013 season, with a range of time of the event of maximum brightness ($T_{\rm max}$) from 2456342.688 to 2456533.251 and 21 events from the 2014 season with $T_{\rm max}$ range from 2456726.323 2456914.471 (Table~\ref{tab1}). The $T_{\rm max}$ is specified in $HJD$ (Heliocentric Julian Date). To emphasize, we only used $I$-band photometric data from the EWS-OGLE database. Every selected event meets three microlensing light curve model parameters: the maximum amplification ($A_{\rm max}$) $\leq$ 6.5 (intensity units), the magnitude amplification ($D_{\rm mag}$) $<$ 1.9 magnitude, and the base magnitude of the source in $I$-band ($I_0$)$ \geq$ 19.5 magnitude. 

\begin{table}
\bc
\begin{minipage}[]{120mm}
\caption{40 Selected Microlensing Events from 2013-2014 seasons.} 
\label{tab1}\end{minipage}
\setlength{\tabcolsep}{5pt}
\small
 \begin{tabular}{ccc}
  \hline\noalign{\smallskip} 
Event & $Tmax$ & $I_0$  \\
          & ($HJD$) & ($mag$) \\
  \hline\noalign{\smallskip}
OB130007 & 2456342.688 & 21.255\\
OB130053 & 2456340.052 & 20.316\\
OB130123 & 2456334.944 & 19.739\\
OB130131 & 2456359.620 & 19.617\\
OB130164 & 2456360.641 & 20.130\\
OB130386 & 2456378.117 & 20.036\\
OB130480 & 2456393.939 & 21.957\\
OB130499 & 2456396.615 & 21.576\\
OB130513 & 2456407.534 & 19.846\\
OB130553 & 2456405.921 & 20.037\\
OB130591 & 2456407.099 & 20.622\\
OB130671 & 2456431.497 & 21.214\\
OB130708 & 2456433.665 & 20.419\\
OB130871 & 2456448.971 & 20.739\\
OB131029 & 2456457.083 & 20.829\\
OB131240 & 2456489.517 & 20.170\\
OB131441 & 2456394.300 & 20.387\\
OB131543 & 2456508.784 & 22.172\\
OB131785 & 2456533.251 & 20.356\\
OB140042 & 2456726.323 & 19.662\\
OB140300 & 2456733.427 & 19.985\\
OB140326 & 2456693.378 & 20.088\\
OB140565 & 2456761.503 & 20.292\\
OB140575 & 2456765.740 & 21.055\\
OB140585 & 2456764.259 & 20.262\\
OB140655 & 2456771.968 & 20.072\\
OB140781 & 2456785.642 & 21.040\\
OB140847 & 2456792.997 & 20.481\\
OB140885 & 2456797.602 & 20.410\\
OB141033 & 2456818.710 & 19.790\\
OB141106 & 2456820.626 & 20.950\\
OB141117 & 2456823.992 & 21.082\\
OB141148 & 2456830.665 & 20.052\\
OB141202 & 2456832.262 & 22.056\\
OB141229 & 2456833.003 & 20.022\\
OB141283 & 2456840.671 & 21.021\\
OB141605 & 2456878.619 & 20.299\\
OB141609 & 2456880.637 & 19.831\\
OB141691 & 2456887.918 & 20.910\\
OB141885 & 2456914.471 & 21.616\\

  \noalign{\smallskip}\hline
\end{tabular}
\ec
\tablecomments{0.86\textwidth}
{accessed OGLE site (http://ogle.astrouw.edu.pl/ogle4/ews/ews.htm) at June 25th, 2015.}
\end{table}

\section{Result}
\label{sec:Result}
\subsection{The flux simulation}
\begin{figure}
   \centering
   \includegraphics[width=10 cm, angle=0]{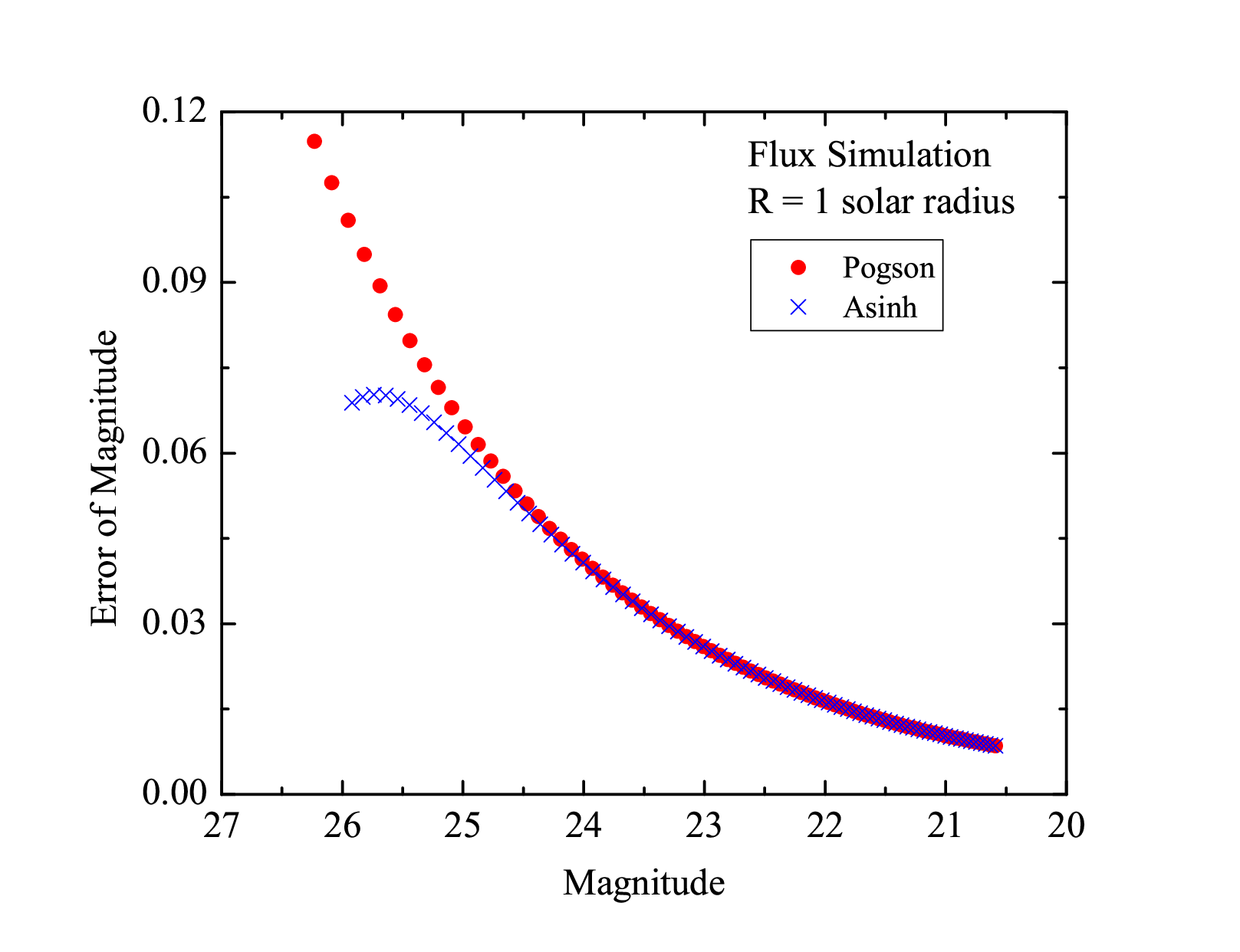}
   \caption{The plot of Error of Magnitude vs  Magnitude for 1 $R_\odot$ for stellar temperatures ranges from 3000 to 11000 $K$ (Ibrahim et al.~\cite{ibrahim15}).}
   \label{Fig2}
   \end{figure}

\begin{figure}
   \centering
   \includegraphics[width=10 cm, angle=0]{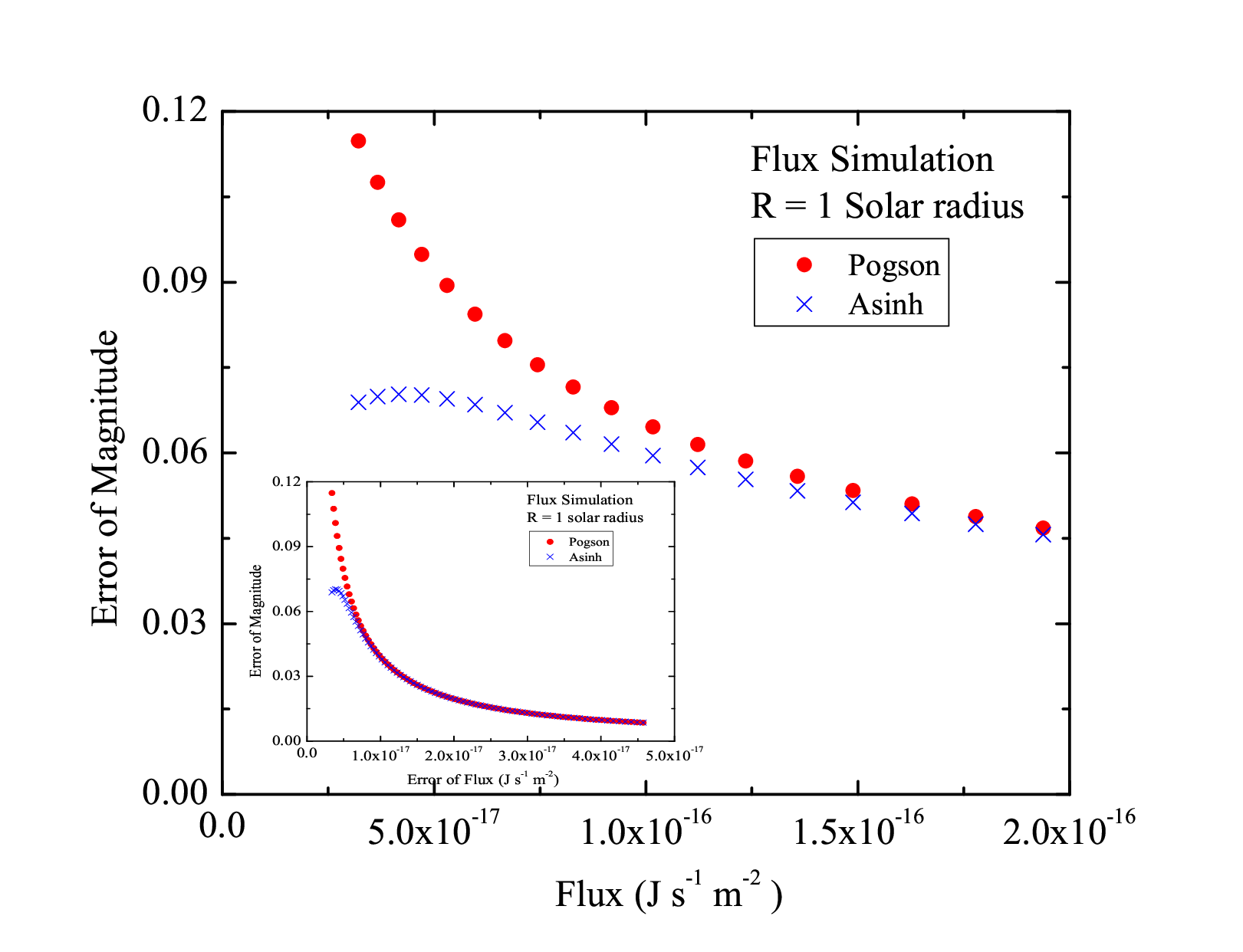}
   \caption{The plot of Error of Magnitude vs Flux for 1 $R_\odot$ for stellar temperature range from 3000 to 4700 $K$. The Flux is in Joules second$^{-1}$ meter$^{-2}$ (J s$^{-1}$ m$^{-2}$). The inset picture shows the same plot but for stellar temperatures ranges between 3000 - 11000 $K$. (Ibrahim et al.~\cite{ibrahim15}).}
   \label{Fig3}
   \end{figure}

\begin{figure}
   \centering
   \includegraphics[width=10.0cm, angle=0]{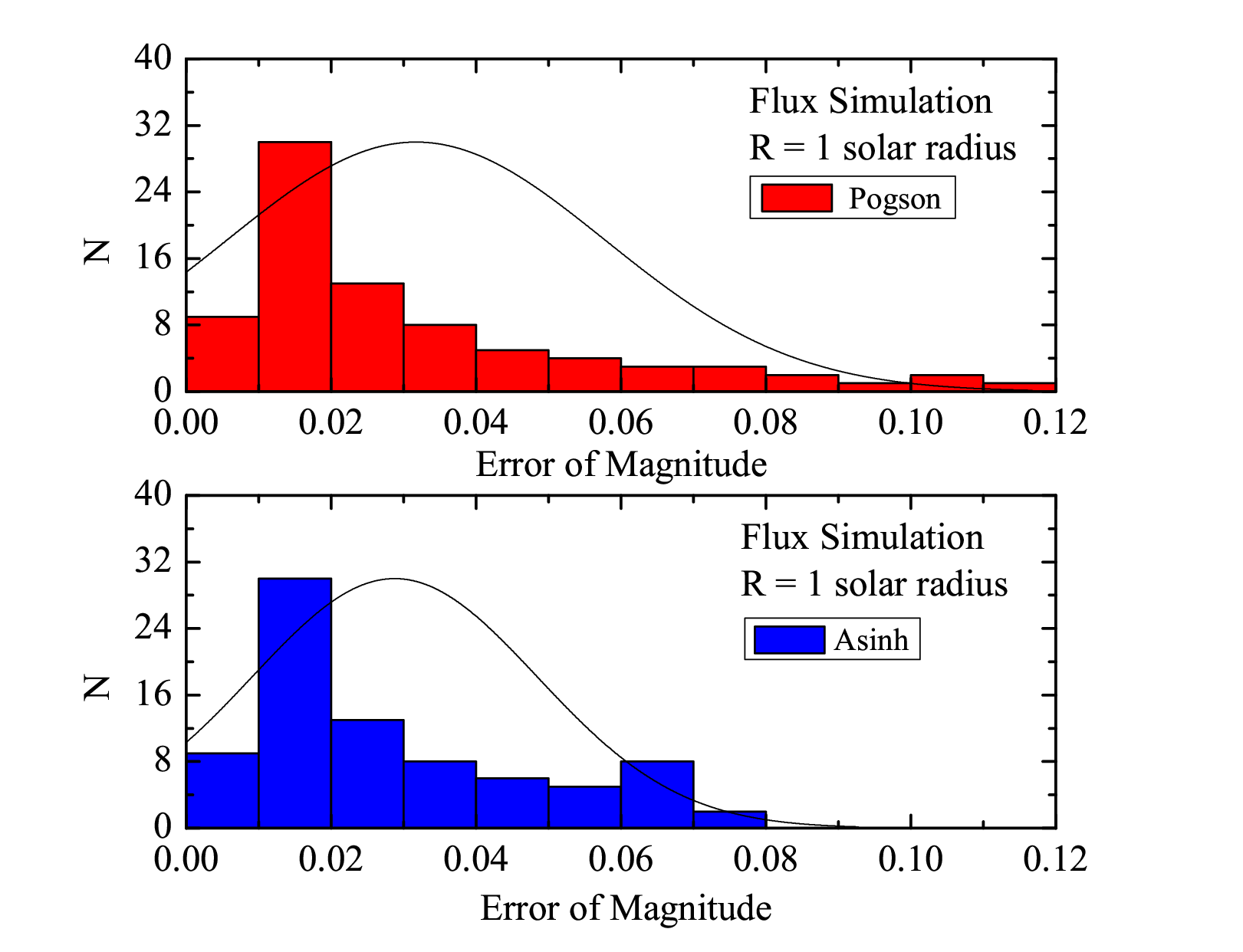}
   \caption{ The histogram of the Error of Magnitude for Pogson and Asinh (from the flux simulation for 1 $R_\odot$ for stellar temperatures ranges from 3000 to 11000 $K$). The line is for the Normal distribution curve on binned data.}
   \label{Fig4}
   \end{figure}

\begin{figure}
   \centering
   \includegraphics[width=10.0cm, angle=0]{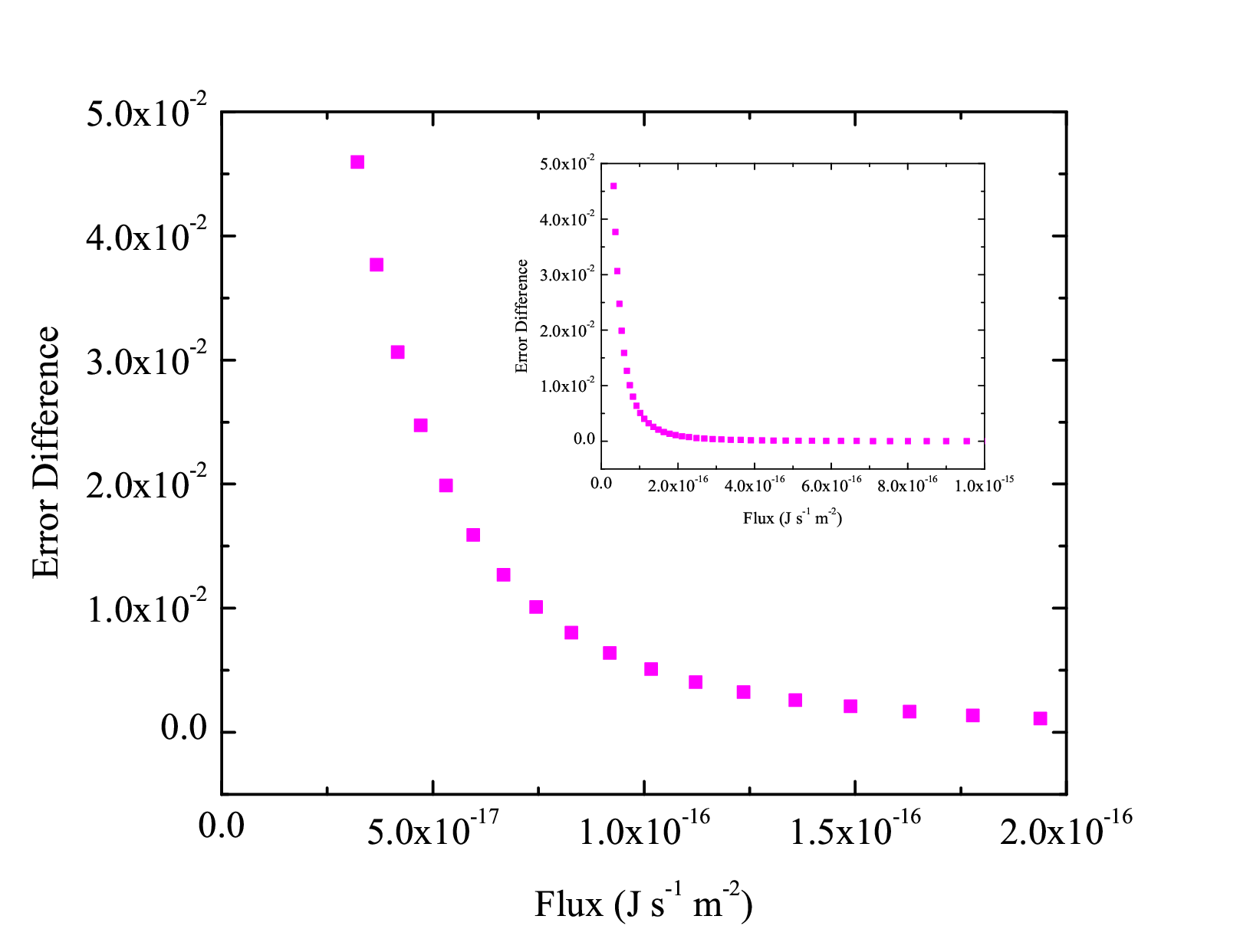}
   \caption{The relationship of the Error Difference (the Pogson Error of Magnitude - the Asinh Error of Magnitude) with Flux for 1 $R_\odot$ for stellar temperature range from 3000 to 4700 $K$. Flux is in J s$^{-1}$ m$^{-2}$. The inset picture shows the same plot but for stellar temperatures ranges between 3000 - 11000 $K$.}
   \label{Fig5}
   \end{figure}

The flux simulation with input star radius of 1 $R_\odot$ resulted in the lowest values of flux. The lowest fluxes were found to produce the largest magnitude values and the largest errors of the flux. In this case, it can be seen that the Asinh formulae gave a smaller error range than the Pogson equations (Figure~\ref{Fig2} and \ref{Fig3}). We also found that the Error of Magnitude histogram for the Asinh formula is narrower than that for the conventional method (Figure~\ref{Fig4}). Statistically, the Asinh formulae gave better flux-to-magnitude transformation. Consequently, we also found that the Error Difference, which was defined as the difference between the Pogson Error of Magnitude and the Asinh Error of Magnitude values (the Pogson Error of Magnitude - the Asinh Error of Magnitude), increased as the flux was decreased (Figure~\ref{Fig5}).

\begin{figure}
   \centering
   \includegraphics[width=10.0cm, angle=0]{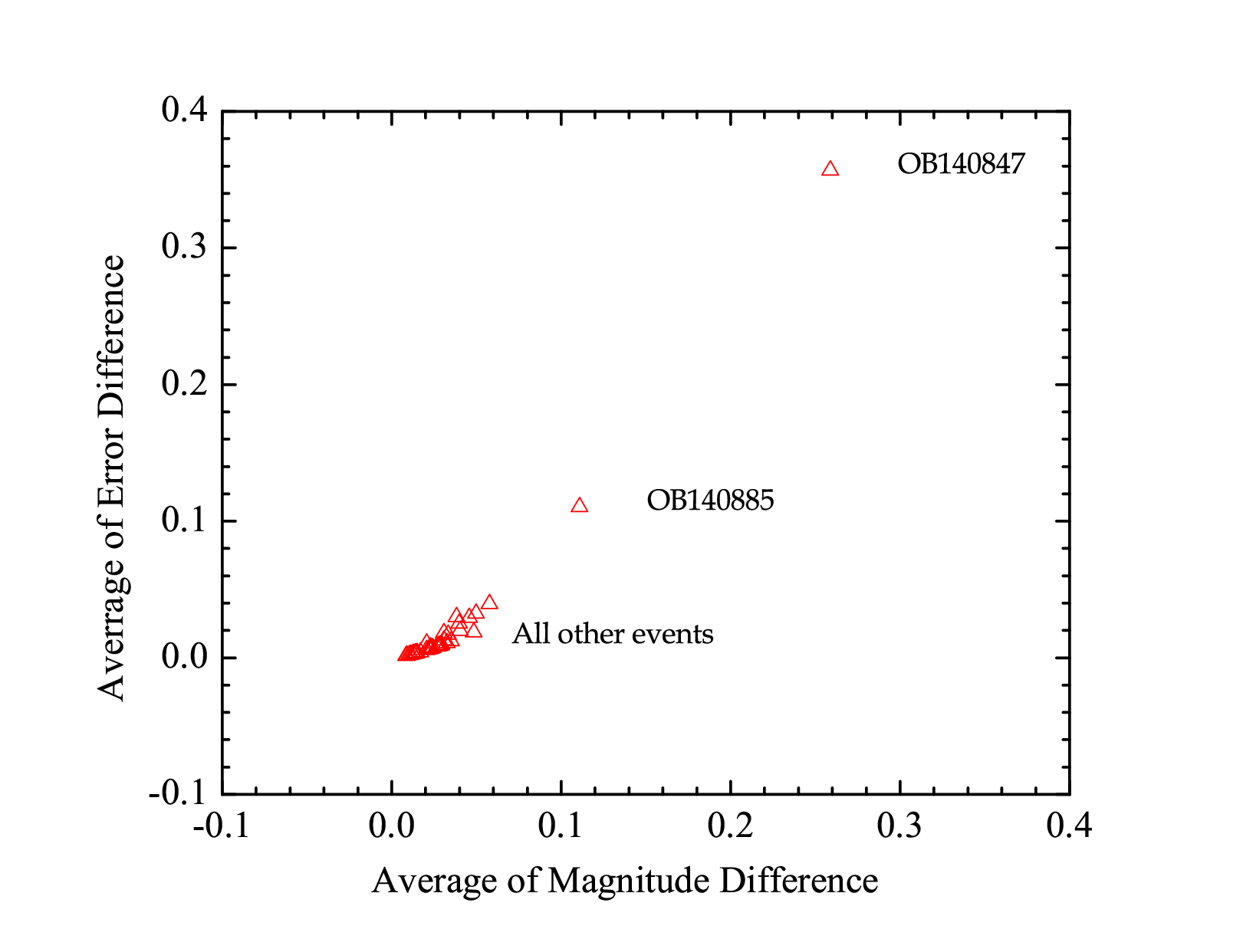}
   \caption{The plot of the average of the Error Difference and the average of the Magnitude Difference for 40 selected events. All the selected events (``All other events'') are grouping in an area with an average of the Magnitude Difference $<$ 0.06 and average of the Error Difference $<$ 0.04  except two events, OB140847 and OB 140885.}
   \label{Fig6}
   \end{figure}
 
\begin{figure}
   \centering
   \includegraphics[width=12.0cm, angle=0]{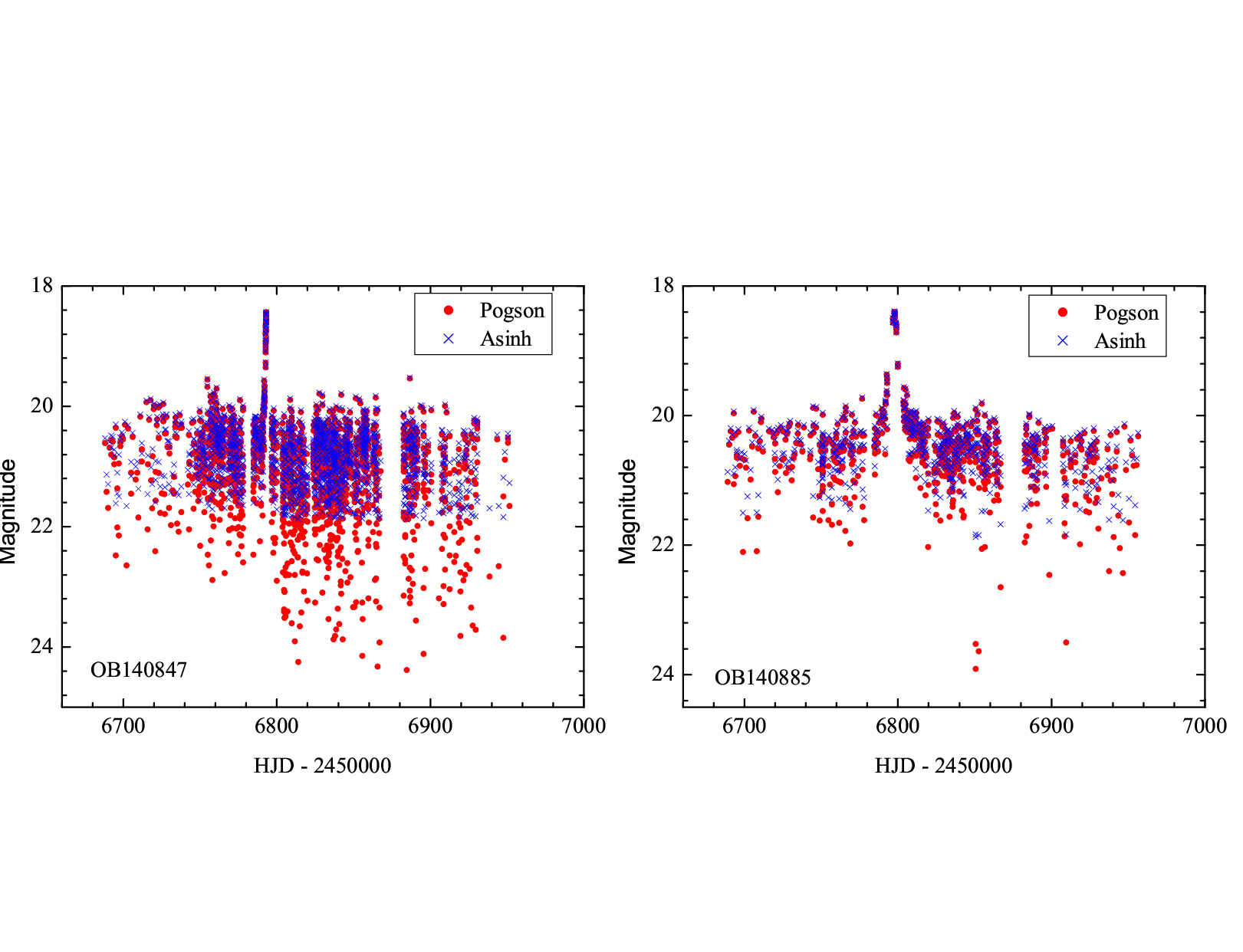}
   \caption{The light curve of OB140847 and OB140885 events. Both events have the largest of the average of the Magnitude Difference and the Error Difference (Table~\ref{tab2}). The red circles are the Pogson magnitude and the blue crosses are the Asinh magnitude. The horizontal axis is (HJD - 2450000) and the vertical axis is the Magnitude.}
   \label{Fig7}
   \end{figure}
  
\begin{figure}
   \centering
   \includegraphics[width=12.0 cm, angle=0]{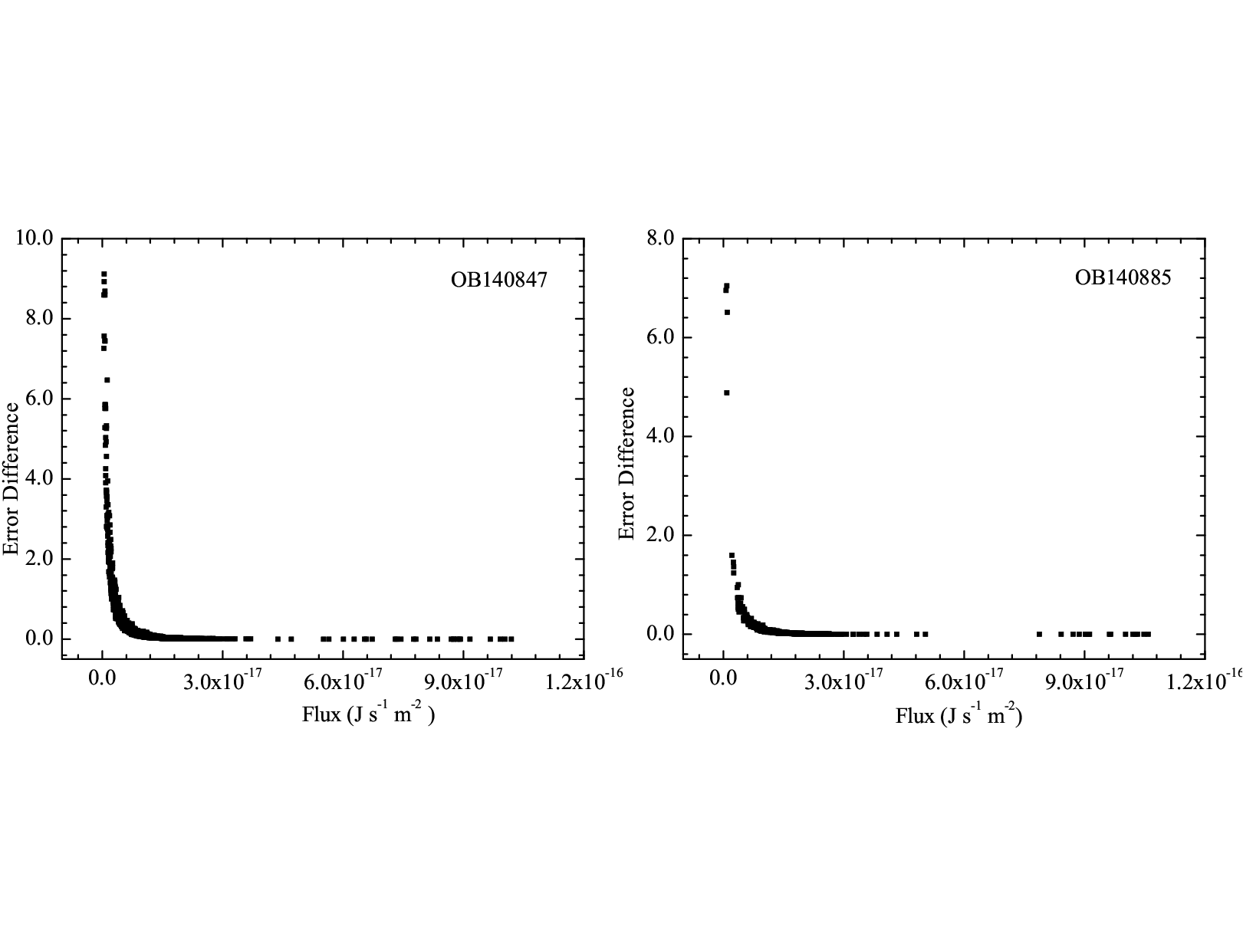}
   \caption{The plot of the Error Difference (the Pogson Error of Magnitude - the Asinh Error of Magnitude) with the Flux for OB140847 and OB140885 events. The horizontal axis is the Flux and the vertical axis is the Error Difference. Flux is in J s$^{-1}$ m$^{-2}$.}
   \label{Fig8}
   \end{figure}
  
\begin{figure}
   \centering
   \includegraphics[width=12.0cm, angle=0]{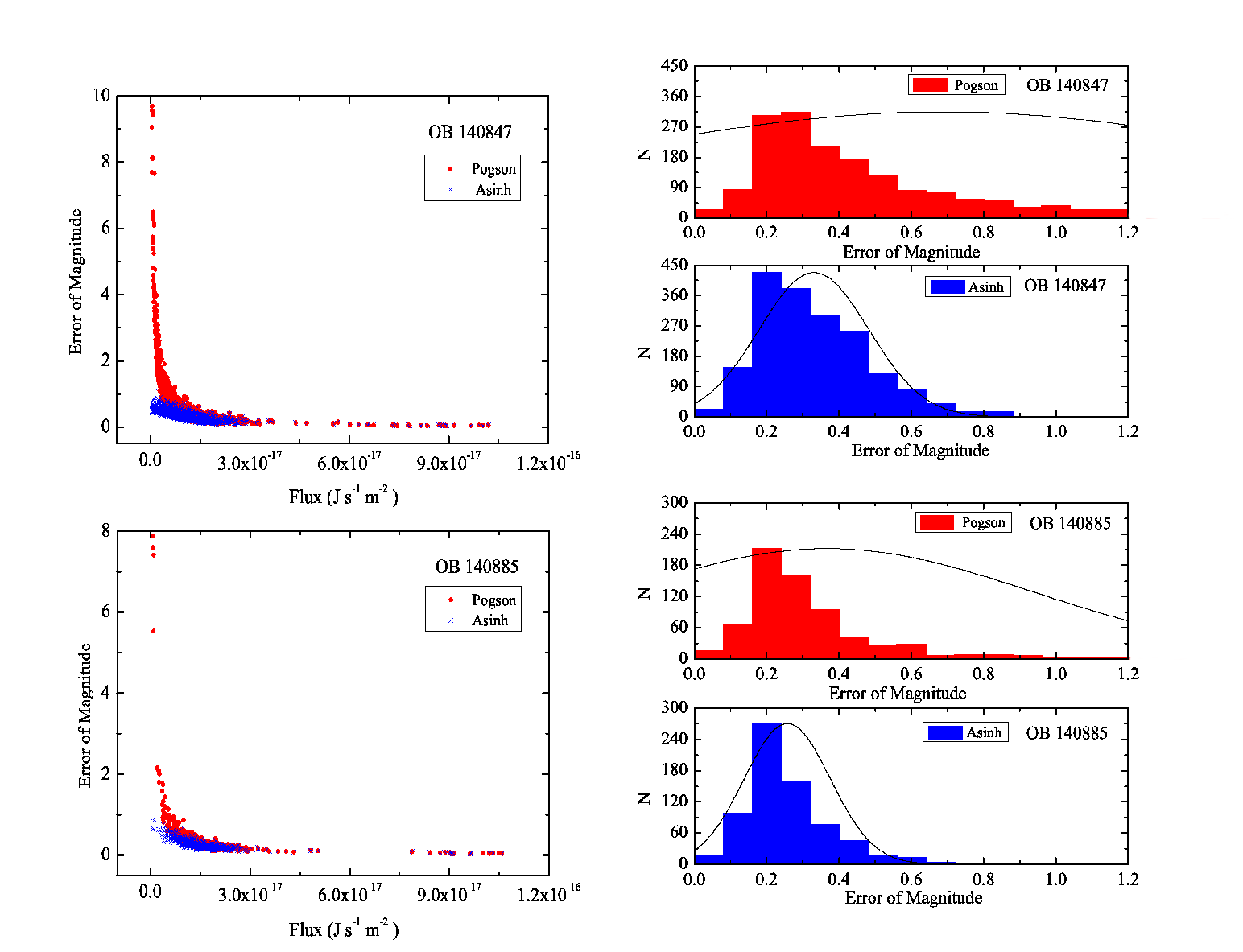}
   \caption{(Left panel) The plot of the Error of Magnitude with their fluxes for OB140847 and OB140885. The range of the Pogson Error of Magnitude is 0 to 10 for OB140847 and 0 to 8 for OB140885. The red circles are the Pogson magnitude and the blue crosses are the Asinh magnitude. (Right panel) The histograms of the Error of Magnitude for OB140847 and OB140885. The bin sizes for all the histograms are 0.08. The ranges of the Pogson Error of Magnitude are 0 to 10 for OB140847 and 0 to 8 for OB140885. The range of the Asinh Error of Magnitude for both events is 0 to 1.2. There are 125 bins for the Pogson and 15 bins for the Asinh. For the purposes of display in this figure, the ranges of Error of Magnitude are set to 0 to 1.2 for each magnitude. The line that overlays the histogram is for the Normal distribution curve on binned data. The histograms of the Pogson Error of Magnitude are wider than the Asinh Error of Magnitude and no closer to the Gaussian than the Asinh.}
   \label{Fig9}
   \end{figure}
    
\subsection{The photometry data from the selected events}
First, the photometry data of the selected events are transformed to the flux and its error. After that, we used equations (5) and (6) to calculate the Asinh magnitude and their errors for all selected events. Because the events occur over time, there is not a single magnitude value or error value but an average over time. For analysis of a selected events, we calculate the average of magnitudes and the average of the Error Difference. The average of the Magnitude Difference has a range from $8.556\times10^{-3}$ to $2.589\times 10^{-1}$  magnitude and the average of the Error Difference has a range from $1.495\times10^{-3}$ to $3.567\times10^{-1}$ magnitude (Figure~\ref{Fig6}).

The data of the differences are presented in Table~\ref{tab2}. From the figure, we can see all of the selected events are grouping in an area with an average of Magnitude Difference $<$ 0.06 and an average of Error Difference $<$ 0.04  except two events, OB140847 and OB140885. Both of the events have the largest of the differences, the largest of the standard deviations of magnitude, the smallest signal to noise ratios, and the faintest magnitudes among the selected events (Table~\ref{tab2}). Because of these facts, we chose them for further analysis and light curves are presented in Figure~\ref{Fig7}. We could see that the Asinh magnitude has a smaller value than the Pogson magnitude. The range of the Asinh magnitude is about 18.4 to 21.9 for both events. But, the ranges of Pogson magnitude are about 18.4 to 24.4 (OB140847) and 18.4 to 24.0 (OB140885). Also, the Asinh magnitude gives a smaller Error of Magnitude than the Pogson magnitude. The largest average of the Error Difference is 0.357 from one event: OB140847 and the smallest is 0.002 from four events: OB130123, OB130131, OB130671, and OB131691 (Table~\ref{tab2}). These facts can be seen more firmly in the examples given in Figure~\ref{Fig8}, which shows the relationship of the Error Difference and the Flux from both methods. The results are similar to the results from subsection 4.1. Further, we compare graphs of the Magnitude Difference versus Flux with histograms both magnitude. These are shown in Figure~\ref{Fig9}. From the figure, we see that the histograms of the error from the Asinh are narrower than the logarithmic (Pogson). 

\begin{figure}
   \centering
   \includegraphics[width=12.0cm, angle=0]{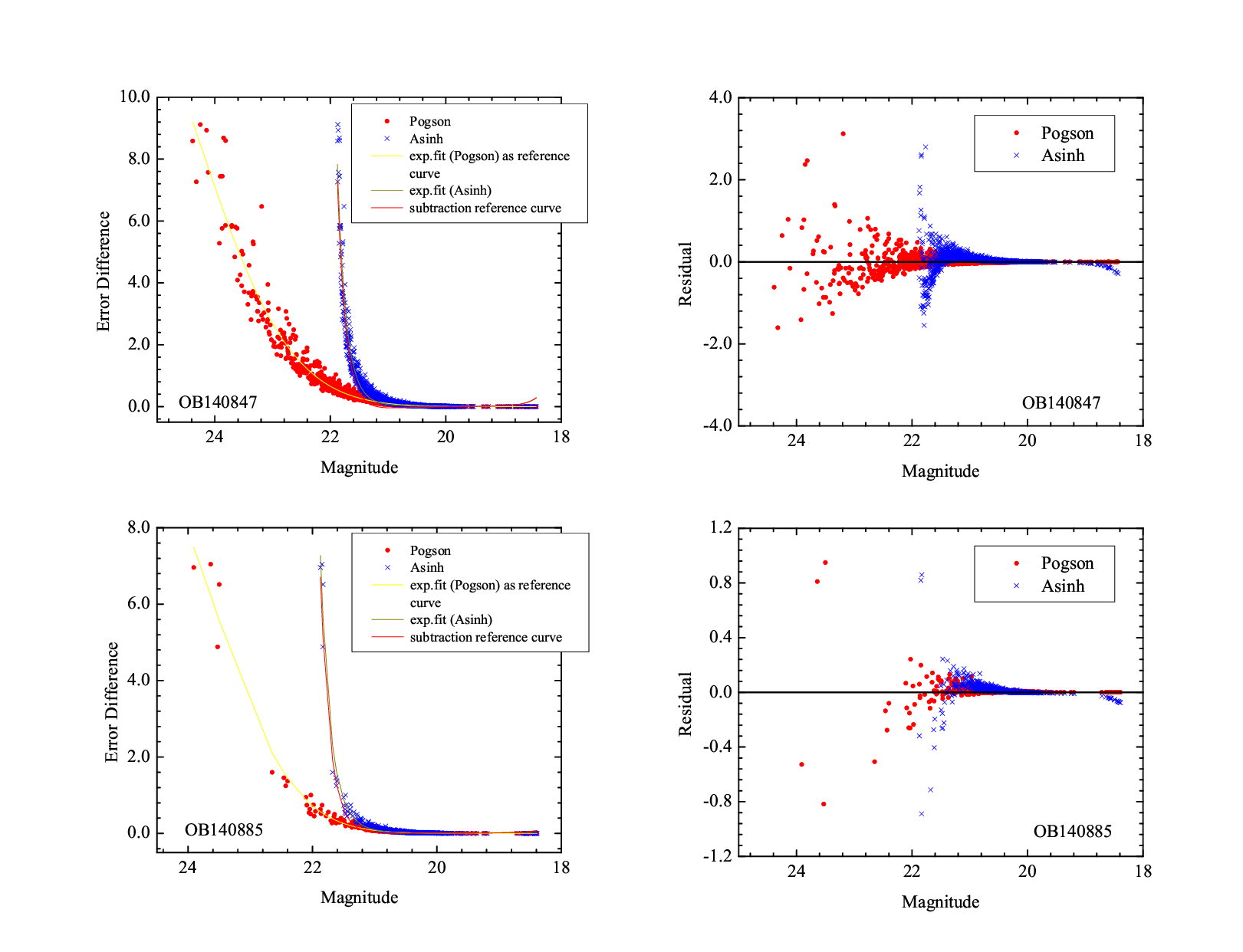}
   \caption{The plot of the Error Difference (the Pogson Error of Magnitude - the Asinh Error of Magnitude) $vs$ the Magnitude (left panel) and the fitting residual plots (right panel) for OB140847 and OB140885 events. The red circles for the Pogson, the blue crosses for the Asinh. The yellow line is exponential fit of the Error Difference to the Pogson Magnitude and the green line is exponential fit of the Error Difference to the Asinh Magnitude. The function in equation (8) is using for the exponential fitting. Coefficient of determination of exponential fit for the Pogson are 0.9555 (OB140847) and 0.962 (OB140885) and for the Asinh are 0.926 (OB140847) and 0.922 (OB140885). The exponential fit of the Error Difference to Pogson Magnitude are set as the reference data set/reference curve. The exponential fit of the Asinh will be subtracted by the reference data set/curve. Before the subtraction, the two data sets would be interpolated or extrapolated. The red line is the result of the subtraction. From the result, what magnitude has an Error Difference of about 0.001 and 0.0001, could be determined and called as ``magnitude limits''.}
   \label{Fig10}
   \end{figure}

We found another interesting result from plotting the Error Difference with Magnitude (Figure ~\ref{Fig10}), there is a tendency that the Error Difference will grow rapidly for small magnitude. We want to know if there is a equation that can describes the Difference Error as a function of magnitude. 

To find the equation, we did the curve fitting to the data set. On the basis of the fact that there is a tendency that the Error Difference will expand exponentially for the small magnitude, the authors choose an exponential function (Ibrahim et al.~\cite{ibrahim15}) to relate Error Difference ($\delta$) with each magnitude ($m$). The relation is given in equation (8).

\begin{equation}
	\delta = e^{(a+bm+cm^{2})}
\label{eq:curvefit}
\end{equation}
     where 
     $\delta$ is Error Difference, 
     $m$ is magnitude, 
     $a$ is a constant of the function, and $b$, $c$ are coefficients of the function.

Later, we found that the coefficient of determination ($R^{2}$) of curve fitting is always greater than 0.72 (Table~\ref{tab3}). The smallest coefficient of determination is 0.725 (OB141691) and the largest is 0.988 (OB130553). The examples of the model fitting and their residuals are presented in Figure~\ref{Fig10}. Given the goodness of fit, we could say that the Error Differences are dependent on their magnitude.

\begin{table}
\bc
\begin{minipage}[]{120mm}
\caption{Average sky levels, S/N ratios, statistical data, and groupings 
based on hypothetical population test for 40 selected events.}
\label{tab2}\end{minipage}
\setlength{\tabcolsep}{2.5pt}
\small
 \begin{tabular}{ccccccccc}
  \hline\noalign{\smallskip}
Event &  Sky Level & $S/N$ & The number & Minimum & Std.Dev of & Average of & Average of& Grouping based  \\
          & & & of data & Magnitude & of Magnitude & Magnitude Difference& Error Difference & Hyp.Population Test\\
  \hline\noalign{\smallskip}
OB130007 & 495.400 & 6.63 & 2675 & 21.198 & 0.258 & 0.033 & 0.011 & I \\
OB130164 & 590.329 & 7.14 & 4779 & 21.208 & 0.289 &  0.029 & 0.009 & I \\
OB130386 & 593.107 & 7.55 & 877 & 22.852 & 0.377 & 0.033 & 0.017 & I \\
OB130480 & 524.486 & 6.99 & 5001 & 21.088 & 0.217 & 0.029 & 0.009 & I \\
OB130499 & 630.454 & 8.60 & 2305 & 23.185 & 0.450 & 0.031 & 0.018 & I \\
OB130591 & 426.797 & 8.32 & 1705 & 24.142 & 0.528 & 0.038 & 0.030 & I \\
OB131029 & 620.714 & 6.97 & 3615 & 23.171 & 0.462 & 0.046 & 0.029 & I \\
OB131543 & 640.079 & 8.43 & 5320 & 20.882 & 0.234 & 0.021 & 0.006 & I \\
OB140042 & 623.243 & 10.51 & 10532 & 20.649 & 0.240 & 0.014 & 0.003 & I \\
OB140326 & 732.749 & 5.99 & 2181 & 24.140 & 0.388 & 0.050 & 0.032 & I \\
OB140575 & 601.018 & 8.05 & 3738 & 20.822 & 0.270 & 0.023 & 0.007 & I \\
OB140655 & 732.144 & 6.75 & 2187 & 23.100 & 0.384 & 0.040 & 0.020 & I \\
OB140781 & 639.498 & 6.38 & 2176 & 22.940 & 0.498 & 0.058 & 0.039 & I \\
OB140847 & 614.790 & 3.33 & 1816 & 24.385 & 0.815 & 0.259 & 0.357 & I \\
OB140885 & 575.835 & 4.69 & 698 & 23.909 & 0.617 & 0.111 & 0.110 & I\\
OB141033 & 577.601 & 9.29 & 9376 & 20.659 & 0.263 & 0.017 & 0.004 & I \\
OB141117 & 531.100 & 7.72 & 787 & 22.360 & 0.369 & 0.031 & 0.013 & I \\
OB141202 & 554.643 & 6.90 & 9729 & 21.107 & 0.225 & 0.029 & 0.009 & I \\
OB141283 & 538.156 & 8.19 & 5971 & 20.952 & 0.256 & 0.022 & 0.006 & I \\
OB141885 & 566.310 & 8.13 & 9001 & 20.921 & 0.268 & 0.023 & 0.007 & I \\

OB130053 & 484.447 & 6.89 & 778 & 21.162 & 0.327 & 0.035 & 0.012 & II \\
OB130123 & 421.165 & 12.46 & 3212 & 20.550 & 0.167 & 0.009 & 0.002 & II \\
OB130131 & 568.990 & 11.39 & 2902 & 20.570 & 0.223 & 0.011 &0.002 & II \\
OB140585 & 437.112 &  7.25 & 1139 & 21.123 & 0.282 & 0.028 & 0.009 & II \\
OB141691 & 533.928 & 12.78 & 4193 & 20.402 & 0.151 & 0.009 & 0.002 & II\\

OB130871 & 629.174 & 11.77 & 2300 & 22.855 & 0.551 & 0.021 & 0.011 & III \\

OB130513 & 432.907 & 12.97 & 2367 &20.964 & 0.486 & 0.013 & 0.003 & IV \\
OB130553 & 534.685 & 10.04 & 89 & 22.145 & 0.730 & 0.040 &0.025 & IV \\
OB130671 & 561.398 & 11.91 & 996 &20.274 & 0.211 &0.010 &0.002 & IV \\
OB130708 & 449.479 & 11.82 & 326 &20.615 & 0.352 & 0.014 &0.003 & IV \\
OB131240 & 414.798 &  9.27  & 865 &21.586 & 0.410 & 0.023 &0.008 & IV \\
OB131441 & 395.931 &  7.59  & 144 & 20.945 & 0.303 & 0.028 &0.009 & IV \\
OB131785 & 369.971 & 7.67 & 172 & 20.964 & 0.329 & 0.027 & 0.008 & IV \\
OB140300 & 524.202 & 10.32 & 258 & 20.967 & 0.525 & 0.022 & 0.007 & IV \\
OB140565 & 413.912 &  7.66 & 467 &21.092 & 0.337 & 0.028 & 0.009 & IV \\
OB141106 & 261.888 &  5.38 & 188 &21.526 & 0.320 & 0.048 & 0.019 & IV \\
OB141148 & 392.784 & 10.63 & 268 &20.506 & 0.331 &0.015 & 0.003 & IV \\
OB141229 & 399.928 & 11.98 & 152 & 20.711 & 0.483 & 0.014 &0.003 & IV \\
OB141605 & 483.014 &  7.82  & 510 &20.938 & 0.272 & 0.024 & 0.007 & IV \\
OB141609 & 442.155 &  10.88 & 472 &20.374 & 0.242 & 0.013 &0.003 & IV\\
  \noalign{\smallskip}\hline
\end{tabular}
\ec
\tablecomments{0.86\textwidth}
{\\
I. reject null hypothesis for both test\\
II. reject null hypothesis for the t- test only\\
III. reject null hypothesis for the variances test only\\
IV. accept null hypothesis for both test}
\end{table}

\begin{table}
\bc
\begin{minipage}[]{100mm}
\caption[]{Coefficients of determination ($R^{2}$) of fitting for the function $\delta = e^{(a+bm+cm^{2})}$  and magnitude limits for two values of $|\Delta \delta|$ = 0.001 and 0.0001 magnitude for the selected events.}
\label{tab3}\end{minipage}
\setlength{\tabcolsep}{2.5pt}
\small
 \begin{tabular}{ccccc}
  \hline\noalign{\smallskip}
Event & $R^{2}$ (the Pogson set data) &  $R^{2}$ (the Asinh set data) & $|\Delta \delta|= 0.001$ & $|\Delta \delta|= 0.0001$ \\
  \hline\noalign{\smallskip}
OB130007 & 0.899 & 0.899 & 20.415 & 19.745\\
OB130053 & 0.948 & 0.948 & 20.431 & 19.850\\
OB130123 & 0.785 & 0.785 & 20.382 & 19.947\\
OB130131 & 0.876 & 0.876 & 20.241 & 19.776\\
OB130164 & 0.907 & 0.907 & 20.401 & 19.776\\
OB130386 & 0.943 & 0.944 & 20.282 & 19.968\\
OB130480 & 0.828 & 0.828 & 20.387 & 19.775\\
OB130499 & 0.963 & 0.958 & 20.440 & 18.589\\
OB130513 & 0.950 & 0.949 & 20.114 & 19.669\\
OB130553 & 0.988 & 0.988 & 20.335 & 18.318\\
OB130591 & 0.973 & 0.957 & 20.841 & 20.456\\
OB130671 & 0.823 & 0.823 & 20.539 & 19.849\\
OB130708 & 0.920 & 0.920 & 20.399 & 19.941\\
OB130871 & 0.942 & 0.942 & 20.448 & 18.668\\
OB131029 & 0.949 & 0.950 & 20.245 & 18.390\\
OB131240 & 0.919 & 0.920 & 20.478 & 19.932\\
OB131441 & 0.908 & 0.908 & 20.548 & 20.087\\
OB131543 & 0.854 & 0.854 & 20.309 & 19.823\\
OB131785 & 0.932 & 0.932 & 20.533 & 20.053\\
OB140042 & 0.900 & 0.900 & 20.177 & 19.722\\
OB140300 & 0.935 & 0.935 & 20.285 & 19.863\\
OB140326 & 0.949 & 0.922 & 18.951 & NA\\
OB140565 & 0.935 & 0.935 & 20.504 & 19.997\\
OB140575 & 0.891 & 0.891 & 20.228 & 19.721\\
OB140585 & 0.932 & 0.932 & 20.475 & 19.970\\
OB140655 & 0.937 & 0.938 & 20.261 & 17.357\\
OB140781 & 0.946 & 0.948 & 19.894 & 18.524\\
OB140847 & 0.955 & 0.926 & 20.096 & 19.947\\
OB140885 & 0.962 & 0.922 & 20.577 & 20.082\\
OB141033 & 0.904 & 0.904 & 20.231 & 19.753\\
OB141106 & 0.949 & 0.949 & 20.780 & 20.327\\
OB141117 & 0.925 & 0.928 & 20.424 & 19.082\\
OB141148 & 0.910 & 0.910 & 20.501 & 20.017\\
OB141202 & 0.860 & 0.861 & 20.416 & 19.825\\
OB141229 & 0.949 & 0.949 & 20.505 & 20.037\\
OB141283 & 0.873 & 0.874 & 20.357 & 19.849\\
OB141605 & 0.911 & 0.911 & 20.542 & 20.076\\
OB141609 & 0.887 & 0.887 & 20.357 & 19.887\\
OB141691 & 0.725 & 0.725 & 20.137 & 19.794\\
OB141885 & 0.899 & 0.899 & 20.253 & 19.717\\
  \noalign{\smallskip}\hline
\end{tabular}
\ec
\end{table}

We set the data from the exponential fit of the Error Difference (the Pogson Error of Magnitude - the Asinh Error of Magnitude) to the Pogson Magnitude as the reference data set/reference curve. We subtracted the data of the exponential fit of the Error Difference to the Asinh magnitude by the reference data set/curve. Before the subtraction, the two data sets would be interpolated or extrapolated. Later, we calculate the result of subtract reference curve for each event (in magnitudes). Then, we can determine, from the result of subtraction, what magnitude has an Error Difference of about 0.001 and 0.0001 magnitude. We called them as `magnitude limits'.

To distinguish from the Error Difference (the Pogson Error of Magnitude - the Asinh Error of Magnitude) $\delta$ that is used for exponential fitting of the Error Difference to each magnitude (the Pogson or the Asinh magnitude), we write the result of the subtraction of the exponential fit of the Error Difference to the Pogson Magnitude by the exponential fit of the Error Difference to the Asinh Magnitude as $|\Delta \delta|$. For all the events, we were able to find the magnitude that produced $|\Delta \delta|$ = 0.001. However, we only found magnitude limits from some events that can produce $|\Delta \delta|$ = 0.0001 (Table~\ref{tab3}). We called these "magnitude limits''. Noting that the current limit of detectable stellar variability is on the order of $10^{-3}$ magnitude, the Error Difference of all events could be significant. We firmly recommended using the Asinh magnitude for EWS-OGLE data when the magnitude is about $20.343$ or fainter.

\section{Discussion and Conclusions}
\label{sect:discussioncoclution}
Generally, from the curve fitting, the results from the conventional method look similar to those of the Asinh method. In order to find out whether the data produced from the two approaches are distinguishable, we carried out some hypothetical population testing to determine if the Pogson dataset and the Asinh dataset come from the same population. We used a two-sample t-test and a variances test. The null hypothesis assumes that the data sets are the same. For the t-test, no mean difference between the data sets would indicates that the null hypothesis is true, while a mean difference would support the alternative hypothesis. For the variances test, no variance difference would indicate that the null hypothesis is true, while a ratio of the variances larger than 1 would support the alternative hypothesis.

After performing the tests, we found that 20 out of the 40 selected events reject the null hypothesis for both tests, and hence we conclude that the difference between the Pogson dataset and the Asinh dataset is significant. We then divided the selected events into four groupings:  Group I contains those events that reject the null hypothesis for both tests. Group II contains those events that reject the null hypothesis for the t-test only. Group III contains those events that reject the null hypothesis for the variances test only, and Group IV contains those events that accept the null hypothesis for both tests. We also found that, in general, the average signal-to-noise ratio (S/N) for the 20 selected events (in Group I) is smaller than that for the other events. The results are presented in Table \ref{tab2} and Figure \ref{Fig13}.
The results of the population tests indicate why a half of the selected events have similarity in the residual plots from the curve fitting.

\begin{figure}
   \centering
   \includegraphics[width=12.0cm, angle=0]{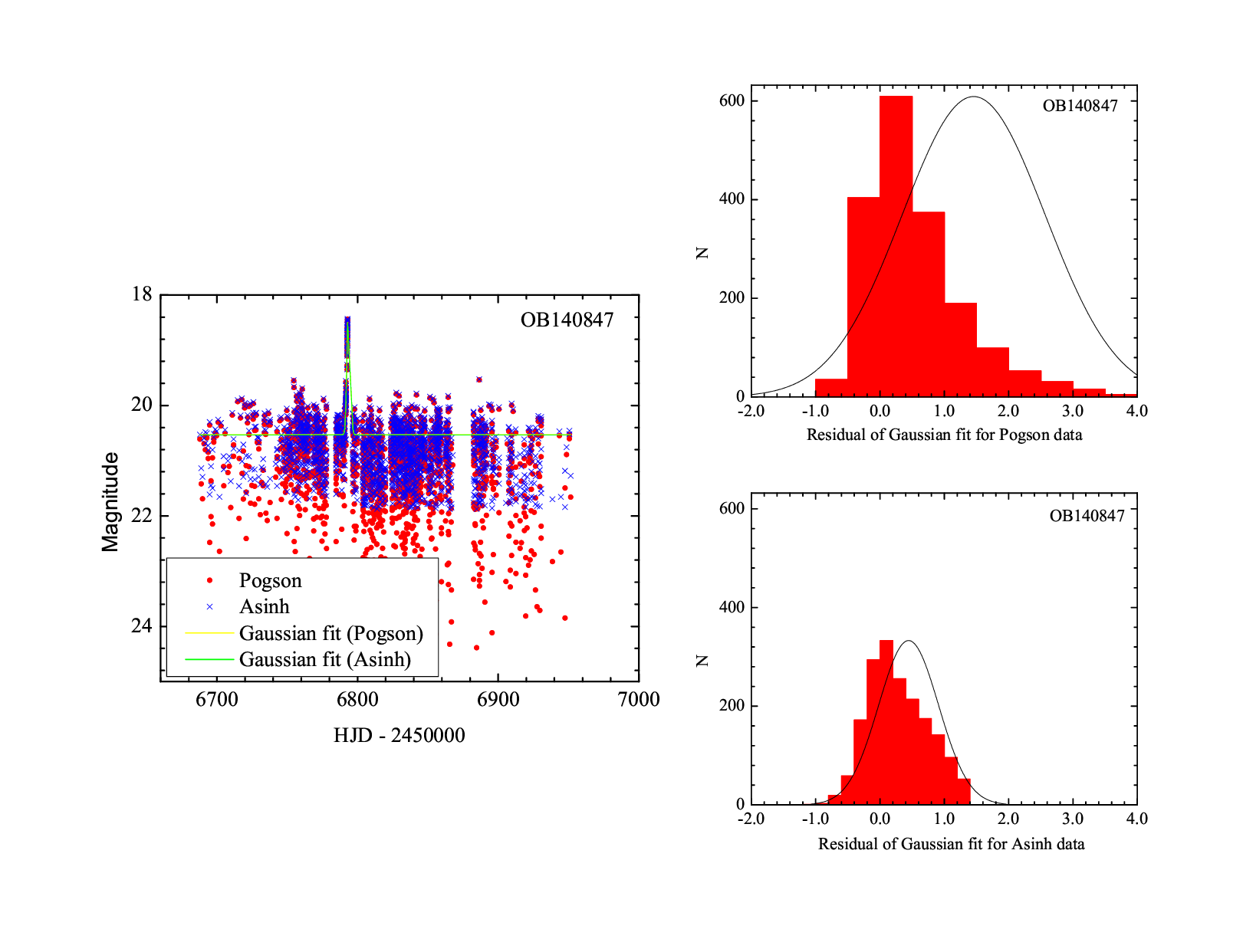}
   \caption{Gaussian fit for the Pogson and the Asinh data and the fitting residual histogram for OB140847. The horizontal axis is for time (HJD - 2450000) and the vertical axis is for event brightness (magnitude). The red circles are for Pogson data, the blue crosses are for the Asinh data, the yellow line is for the Gaussian fitting of the Pogson data, and the green line is for the Gaussian fitting of the Asinh data. The line that overlays the histogram is for the Normal distribution curve on binned data.}
   \label{Fig11}
   \end{figure}
  
\begin{figure}
   \centering
   \includegraphics[width=12.0cm, angle=0]{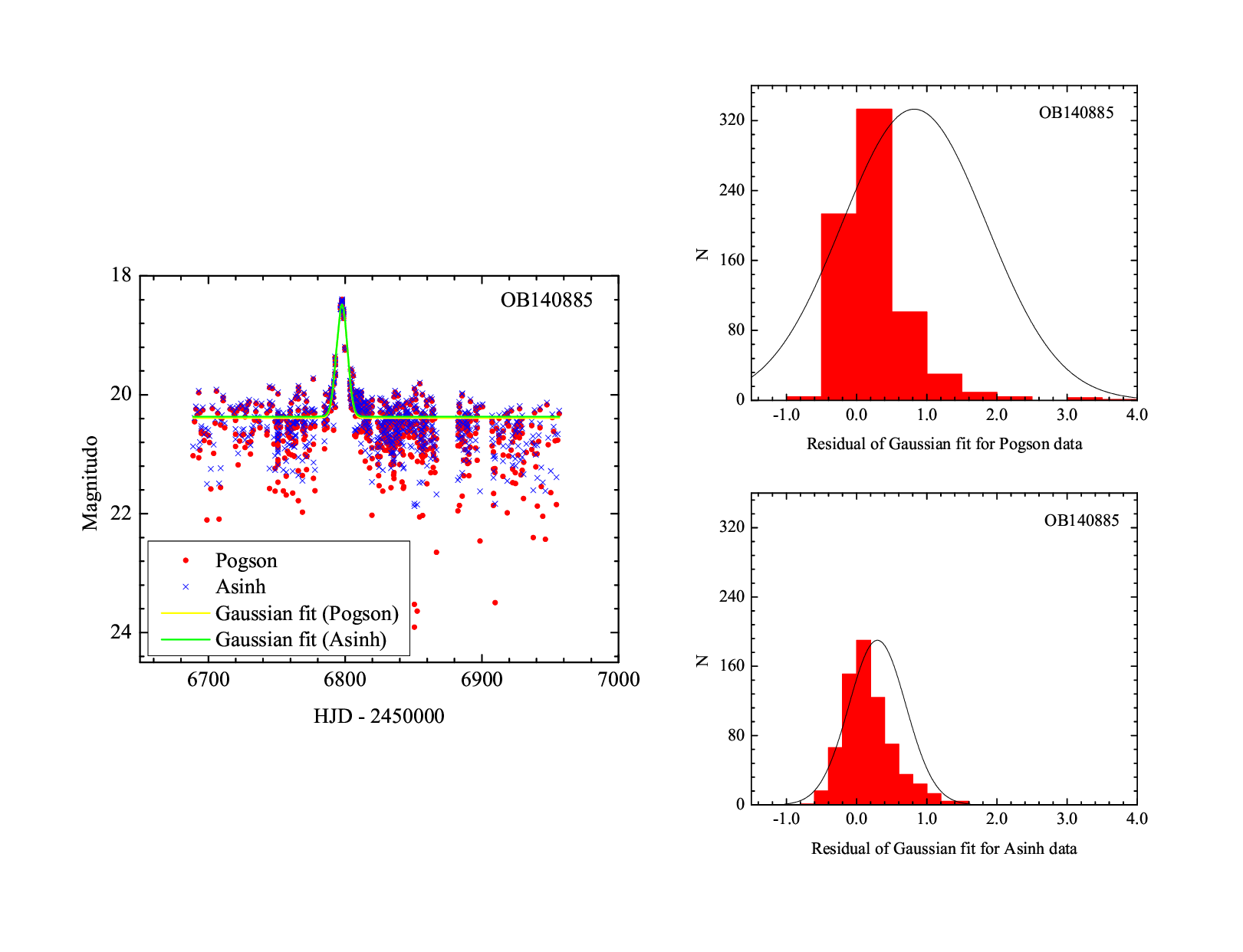}
   \caption{Gaussian fit for the Pogson and the Asinh data and the fitting residual histogram for OB140855. The horizontal axis is for time (HJD - 2450000) and the vertical axis is for event brightness (magnitude). The red circles are for the Pogson data, the blue crosses are for the Asinh data, the yellow line is for the Gaussian fitting of the Pogson data, and the green line is for the Gaussian fitting of the Asinh data. The line that overlays the histogram is for the Normal distribution curve on binned data.}
   \label{Fig12}
   \end{figure}

\begin{figure}
   \centering
   \includegraphics[width=10.0cm, angle=0]{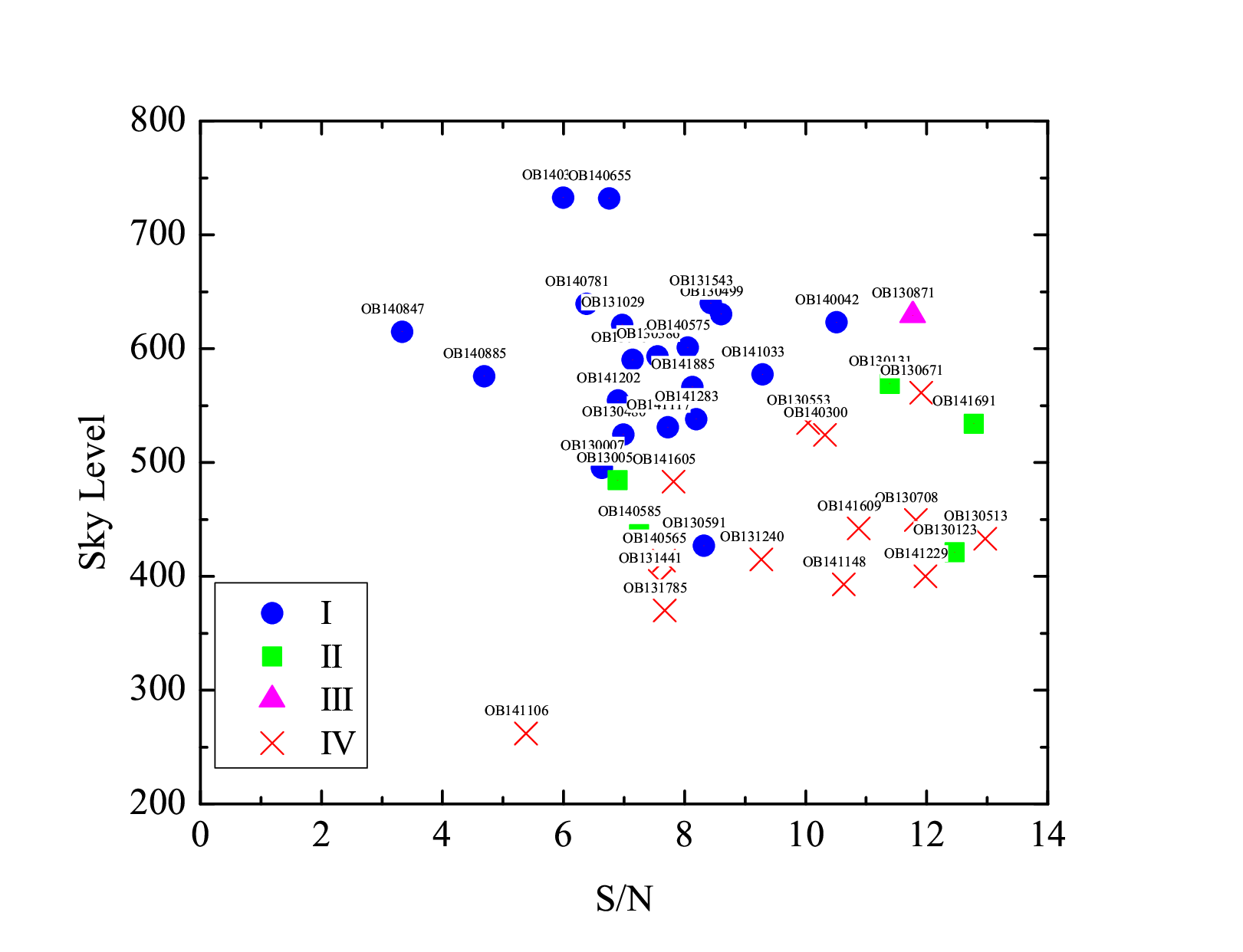}
   \caption{Grouping of hypothetical population test results for all the selected events. The horizontal axis is the average of the signal to noise ratio (S/N) and the vertical axis is the average of the sky level.  Blue circles are for the Group I, green rectangles are for the Group II, magenta triangles are for the Group III, and red crosses are for the Group IV. Two events that have the smallest S/N are OB140847 and OB140885. The Group I have high sky level and the small S/N ratio. The average of S/N and average of sky level are presented in Table~\ref{tab2}.}
   \label{Fig13}
   \end{figure}

By looking at the Error Difference (Error of Pogson - Error of Asinh) $vs$ Flux plots (similar to Figure \ref{Fig8}), it is clear that for larger fluxes, there are no significant differences in magnitude or their error between two methods. However, this is not the case for small fluxes. The Asinh method is found to give a smaller deviation than the Pogson method for the low flux region. The result is similar to that described by Lupton et al.~\cite{lupton99}. We also found that the average of the Magnitude Difference between two methods is in the order of 0.01 magnitude and the Error of Magnitude from the Asinh method is always smaller than that given by the conventional method. The average of Error Difference is in the order of $10^{ -2}$ magnitude. At what limit would we recommend using the Asinh rather than the conventional method? For the instrument used by the OGLE, the limit is 20.343 magnitude. Fainter than that and the Asinh will be better than the conventional method. It is possible to a conduct similar experiment with other instrument systems, to define their magnitude limits. 

We speculate as to why there are 20 events that have a large error and hence for them the Asinh is more appropriate? 

Firstly, they have the faintest magnitude among our selected samples and a high sky level. For example, we can see from the Table~\ref{tab2} that the event OB140847 has the faintest magnitude = 24.385 and sky level = 614.79. After examining photometric data from EWS-OGLE further, it could be seen, in general, that the average sky level during the observation is higher than for other groups (Table~\ref{tab2}). This makes the S/N ratio for these objects smaller than the other groups. The two lowest S/N ratios are 3.33 (OB140847) and 4.69 (OB140885). Another factor that needs to be considered is the number of data. This factor places the OB141106 (the number of data = 188) into Group IV as well as having small S/N. From these, we conclude, there is a combination of several factors such as magnitude, sky level, S/N ratio, and the number of data from the events that results in 20 events requiring use of the Asinh magnitude. 

Secondly, from the result of fitting the magnitude data with the Gaussian function to find the peak of the light curve, there is an indication that the Asinh magnitude is statistically better than the Pogson magnitude for small S/ N data. The parameters of Gaussian fitting for OB140847 and OB140885 data are presented in Table~\ref{tab4}. Other results show that the Asinh magnitude fitting residual histograms are narrower than the Pogson's (Figure~\ref{Fig11} and \ref{Fig12}). In other words, by using the Asinh magnitude, we can determine a more certain base magnitude for each event compared to Pogson magnitude. 

The reduced $\chi^{2}$ = 2.232 (for the Asinh data) is higher than 1.559 (for the Pogson data) from OB140847 and 1.707 (for the Asinh data) is similarly higher than 1.453 (for the Pogson data) from OB140885, but the mean squared values of the regression of the Asinh magnitude were found to be smaller than Pogson magnitude. The results also support the use the Asinh magnitude for small S/N data. The reduced  $\chi^{2}$ values and the mean squared values of the regression for all 40 events are shown in Table~\ref{tab5}. This study provides hope that we could detect microlensing events from a noisier observation when using the Asinh method. If we can retrieve the raw data from the EWS-OGLE database, there is the possibility of discovering other microlensing events from fainter stars. We also hope that our work could be of significant help to the observatories who suffer from an increasingly more light-polluted observation environment, such that they could obtain more valuable signal from fainter stars.

\begin{table}
\bc
\begin{minipage}[]{100mm}
\caption{Parameters of  Gaussian fitting for OB140847 and OB140885.}
\label{tab4}\end{minipage}
\setlength{\tabcolsep}{2pt}
\small
 \begin{tabular}{lcccccccccccc}
  \hline\noalign{\smallskip}
Events&mag&&HJD&&w&&A&&sigma&&FWHM\\
   &Pogson&Asinh&Pogson&Asinh&Pogson&Asinh&Pogson&Asinh&Pogson&Asinh&Pogson&Asinh\\
  \hline\noalign{\smallskip}
OB140847&20.528&20.537&6793.822&6793.830&2.307&2.367&7.878&8.023&1.154&1.184&2.717&2.787 \\
 &$\pm$0.009&$\pm$0.009&$\pm$0.134&$\pm$0.128&$\pm$0.119&$\pm$0.115&$\pm$0.882&$\pm$0.831\\
 OB140885&20.391&20.367&6797.654&6797.666&8.040&8.085&19.216&19.086&4.020&4.042&9.467&9.519\\
 &$\pm$0.012&$\pm$0.011&$\pm$0.097&$\pm$0.104&$\pm$0.176&$\pm$0.184&$\pm$0.464&$\pm$0.475\\
  \noalign{\smallskip}\hline
\end{tabular}
\ec
\end{table}

\begin{table}
\bc
\begin{minipage}[]{100mm}
\caption[]{Reduced $\chi^{2}$ values and Mean squared values of the regression for all 40 events.}
\label{tab5}\end{minipage}
\setlength{\tabcolsep}{2pt}
\small
 \begin{tabular}{lcccc}
  \hline\noalign{\smallskip}
Events&     Reduced $\chi^{2}$&&    Mean Square of the regression&\\
   &Pogson&Asinh&Pogson&Asinh\\
  \hline\noalign{\smallskip}
OB130007&1.932&1.969&275696.298&274779.060\\
OB130053&2.250&2.305&80236.593&79945.576\\
OB130123&2.530&2.536&311715.348&311414.225\\
OB130131&4.052&4.070&277186.127&276847.378\\
OB130164&3.366&3.438&484705.459&483196.018\\
OB130386&3.415&3.536&88346.643&88020.296\\
OB130480&1.634&1.654&510575.351&509087.324\\
OB130499&2.835&2.947&230040.380&229246.443\\
OB130513&3.903&3.928&222143.196&221833.259\\
OB130553&4.324&4.601&8737.589&8695.142\\
OB130591&3.990&4.173&174324.506&173560.100\\
OB130671&2.185&2.190&96111.616&96008.103\\
OB130708&2.947&2.968&32078.768&32029.962\\
OB130871&3.476&3.578&222270.709&221729.266\\
OB131029&2.762&2.924&366430.207&364577.547\\
OB131240&6.485&6.557&86489.845&86273.990\\
OB131441&1.817&1.835&14903.523&14862.271\\
OB131543&2.612&2.639&528581.444&527444.711\\
OB131785&2.204&2.225&17689.260&17641.428\\
OB140042&5.656&5.686&1005100.000&1003590.000\\
OB140300&11.629&11.704&24852.776&24789.817\\
OB140326&2.109&2.201&221161.787&219998.985\\
OB140565&2.691&2.731&47870.199&47731.523\\
OB140575&3.289&3.328&369299.063&368392.245\\
OB140585&2.392&2.426&116803.418&116464.628\\
OB140655&6.227&6.313&219163.844&218219.644\\
OB140781&2.097&2.265&225303.874&223904.455\\
OB140847&1.559&2.232&200588.902&195193.664\\
OB140885&1.453&1.701&73576.156&72716.726\\
OB141033&2.588&2.609&948392.481&946693.061\\
OB141106&1.011&1.028&20487.846&20391.489\\
OB141117&2.293&2.386&80180.387&79913.648\\
OB141148&2.386&2.393&26674.681&26633.978\\
OB141202&1.652&1.677&997291.035&994322.924\\
OB141229&4.304&4.328&14935.737&14913.643\\
OB141283&2.682&2.718&597310.478&595925.593\\
OB141605&1.778&1.798&52394.528&52266.290\\
OB141609&3.250&3.262&46030.650&45969.895\\
OB141691&2.208&2.214&399176.806&398818.338\\
OB141885&2.664&2.716&892879.282&890699.829\\
  \noalign{\smallskip}\hline
\end{tabular}
\ec
\end{table}

\normalem
\begin{acknowledgements}
The writers wish to thank DIKTI Kementerian Riset Teknologi dan Pendidikan Tinggi for the 2014 research grant with contract number: 1062d/I1.C0I/PL/2014.
This paper is a part of Ichsan Ibrahim's dissertation in the Astronomy Graduate Program of Institute Technology of Bandung.
\end{acknowledgements}

\label{lastpage}

\end{document}